\documentclass{aa}     % LaTeX A&A  Standard Fonts

\usepackage{graphicx}
\usepackage{txfonts}

\newcommand{\Mpc}{$h^{-1}$\thinspace Mpc}

\newcommand{\etal}{et al. }
\def\apj{ApJ}
\def\apjl{ApJL} 
\def\apjs{ApJS}

\begin{document}

\title{Clusters and groups of galaxies in the 2dF galaxy redshift survey}

\author {E. Tago\inst{1}, J. Einasto\inst{1}, M. Einasto\inst{1},
  E. Saar\inst{1}}

\authorrunning{E. Tago et al.}

\offprints{E. Tago }

\institute{Tartu Observatory, EE-61602 T\~oravere, Estonia}

\date{ Received   2004 / Accepted ...  }

\titlerunning{2dF clusters}

\abstract{We create a new catalogue of groups and clusters for the 2dF
  GRS final release sample. We show that the variable linking length
  friends-of-friends (FoF) algorithms used so far yield groups with
  sizes that grow systematically with distance from the observer, but
  FoF algorithms with a constant linking length are free from this
  fault.  We apply the FoF algorithm with a constant linking length
  for the 2dF GRS, compare for each group its potential and kinetic
  energies and remove galaxies with excess random velocities.  Our
  sample contains 7657 groups in the Northern part, and 10058 groups
  in the Southern part of the 2dF survey with membership $N_g \geq $
  2.  We analyze selection effects of the catalogue and compare our
  catalogue of groups with other recently published catalogues based
  on the 2dF GRS.  We also estimate the total luminosities of our
  groups, correcting for group members fainter than the observational
  limit of the survey.  The cluster catalogues are available at our
  web-site (\texttt{http://www.aai.ee/$\sim$maret/2dfgr.html}).
 
\keywords{cosmology: observations -- cosmology: large-scale structure
of the Universe; clusters of galaxies}
}

\maketitle

\section{Introduction}

Clusters and groups of galaxies are the basic building blocks of the
Universe.  They have been used for a wide range of studies: properties
of the large-scale structure, galaxy formation and evolution,
environmental studies, studies of dark matter and others.  The
classical cluster catalogues by Abell (\cite{abell}) and Abell, Corwin
and Olowin (\cite{aco}) were constructed by visual inspection of
Palomar plates.  Recently new deep catalogues of galaxies have been
made available, which allow compilation of new catalogues of groups
and clusters of galaxies.  The first catalogue of the new generation
of galaxy groups was the Las Campanas catalogue of groups by Tucker
\etal (\cite{Tucker00}).  The next steps in this field were due to
publication of the SDSS (Sloan Digital Sky Survey) data releases (EDR,
DR1, DR2, DR3) and the 2dF GRS (2 degree Field Galaxy Redshift Survey)
data releases (100K, final). This inspired numerous research teams to
work out more refined cluster finding algorithms and to compile
catalogues of galaxy systems (de Propris \etal \cite{dep02a}; Merchan
\& Zandivarez \cite{mer03}, \cite{mer04}; Bahcall \etal \cite{bac03}; Lee \etal
\cite{lee04}; Eke \etal \cite{eke04}; Einasto \etal \cite{ein04}).  A
large collection of groups of galaxies has been compiled on the basis
of the digitized Palomar Observatory Sky Survey (DPOSS) by Gal \etal
(\cite{gal02}, \cite{gal02}) and Lopes \etal (\cite{lop}).
Intermediate redshift groups of the CNOC2 survey were studied by
Carlberg \etal (\cite{car01}). A recent study of groups of
galaxies was carried out using DEEP2 galaxies (Gerke \etal
\cite{ger04}).

In recent years a number of new group finding algorithms and well
known methods (sometimes modified) have been applied (Kim \etal
\cite{kim02}; Bahcall \etal \cite{bac03}; see a review by Nichol
\cite{nic04}) to these catalogues of galaxies. New methods have been
proposed for finding groups in 2D surveys, using colors of galaxies,
as in the Cut and Enhance method (see, e.g. Goto \etal \cite{goto}),
However, the friend-of-friend (FoF) method is the most frequently
applied for redshift surveys.
 
Kim \etal (\cite{kim02}) and Nichol (\cite{nic04}) have compared
various cluster finding methods.  The use of different methods means
that the resulting catalogues of groups and clusters of galaxies are
rather different, particularly for loose and poor groups.  Recently
Lopes \etal (\cite{lop}) claimed to have obtained agreement between
the two methods they applied (Adaptive Kernel and Voronoi
Tessellation).  However, even a quick inspection of all various group
catalogues reveals both large differences between the lists of these
groups, as well as of their properties. A short review of the problem
is given by Yang \etal (\cite{yang04}).

In addition, in the case of rich clusters there are problems of
substructure in many, if not in a majority of clusters (Burgett \etal
\cite{bur04}; Einasto \etal \cite{ein04}).

Analysis of recent catalogues of groups and clusters of galaxies has
lead us to the conclusion that all the presently available catalogues
have their problems.  In particular, in the only publicly available
carefully compiled catalogue of the 2dF GRS groups by Eke et
al. (\cite{eke04}) that can be checked for systematics, the
characteristic (rms) radii of groups depend on the distance from the
observer (see below).  As an example, this catalogue contains a number
of rich clusters. The dimensions of these clusters in the catalogue
appear to be much larger than expected from Abell counts.  A closer
inspection of these cases demonstrates the existence of substructure
outside the main body of these rich clusters (Einasto \etal
\cite{ein04}).
To improve the cluster catalogue we tried several cluster-finding
algorithms.  After a number of trials we finally selected a series of
procedures discussed below.

{\scriptsize
\begin{table*}
      \caption[]{Data on the 2dF GRS samples and group finding parameters}
         \label{Tab1}
      \[
         \begin{tabular}{ccccccccccccc}
            \hline\hline
            \noalign{\smallskip}
          Sample & $RA$ & $DEC$ & $N_{gal}$ &
      	$N_{groups}$&$N_{single}$&$\Delta$V & $\Delta$R &
	$V_0$ & $R_0$ & $n_{iter}$ \\ 
             & deg & deg &    &
      	& & km/s &  Mpc/h &
	km/s & Mpc/h &  \\ 
            \noalign{\smallskip}
            \hline
            \noalign{\smallskip}

2dF GRS N  & 140...230 & -7...+3 & 78067  & 7657 & 54604 & 750 & 0.3 &
50 & 0.03 & 3 \\  
2dF GRS S  & 320...58  & -37.5...-23 & 106328  & 10058 & 75838 & 750 &
0.3 & 50 & 0.03 & 3 \\  

            \noalign{\smallskip}
            \hline
         \end{tabular}
      \]
   \end{table*}
}

The goal of the present paper is to present a new catalogue of groups
and clusters based on the 2dF Galaxy Redshift Survey. We apply the
simple and well-known friends-of-friends algorithm with a constant
linking length.  In addition we apply a cleaning procedure, based on
the comparison of the potential and kinetic energies of tentative
groups. The data used and the group-finding and cleaning algorithms
are discussed in the next two sections of the paper.  Section 4
describes the group catalogue.  The catalogue is available at the
web-site \texttt{http://www.aai.ee/$\sim$maret/2dfgr.html}.  We also
estimate luminosities of groups and clusters, this is described in
Section 5.  In the last section we discuss the selection effects in
our catalogue and in the 2PIGG catalogue by Eke et al.  We also
compare our groups with groups found by several other
investigators. We use the term ``group'' for all objects in our
catalogue; these include also clusters of galaxies.

\section{The Data}

In this paper we have used the 2dF GRS final release (Colless \etal
\cite{col01}, \cite{col03}) that contains 245591 galaxies. The survey
consists of two main areas in the Northern and Southern hemispheres
within the coordinate patches given in Table~\ref{Tab1}.  As the 2dF
sample becomes very diluted at large distances, we restrict our sample
by a limiting redshift $z=0.2$.  We also apply a lower limit $z \geq
0.009$, as at smaller redshifts the catalogue contains mostly
unclassified objects of the Local Supercluster, and stars.  We do not
restrict our sample by additional magnitude limits. However, we
use the magnitude limit masks, when we find the group luminosity
functions and luminosity weights (Einasto \etal \cite{e03b}).  Our
sample contains 78067 galaxies in the Northern sky and 106328 in the
Southern sky.  However, both the Northern as well as the Southern
areas of the 2dF GRS have rather serrated edges, which can create
serious edge effects. We start from the original galaxy sample, but
restrict later our analysis of groups by more smooth sample edges. The
redshifts were corrected for the motion relative to CMB. For linear
dimensions we use co-moving distances, computed for the matter and
dark energy density parameters 0.3 and 0.7, respectively (see, e.g.,
Mart{\`\i}nez \& Saar (\cite{mar03}).

\section{Cluster finding algorithms}

To search for clusters, the friends-of-friends (FoF) algorithm is
conventionally applied.  This algorithm has two main versions,
suggested by Zeldovich \etal (\cite{zes82}) and by Huchra \& Geller
(\cite{hg82}) (hereafter ZES and HG, respectively).  These algorithms
are essentially identical with one difference: ZES used a constant
linking length to find neighbours, whereas HG applied a variable
linking length $l(r)$ depending on the observed volume density of
galaxies $\rho(r)$ (or the selection factor $f(r)=\rho(r)/\rho(0)$) at
a distance $r$ from the observer.

Several types of scaling of the linking length(s) have been used; see
a summary by Eke et al. (\cite{eke04}); these can be classified by the
cluster models the authors have used, either consciously or
unconsciously. The original Huchra \& Geller scaling, $l\sim
f^{-1/3}$, corresponds to the real-space cluster analysis, where all
distances between galaxies are diluted by the same factor.  In real
observations, we should differentiate between projected distances in
the sky and radial velocity-space distances. The projected distances
scale under dilution as $l_{\mbox{sky}}\sim f^{-1/2}$. The sizes of
observed clusters in velocity space ($\Delta v/H$) are usually about
10 times larger than their radial sizes.  It means that all position
information is practically lost in velocity space, and the linking
length in this space should remain constant, or more exactly, scale by
the usual Doppler broadening, $l_v\sim(1+z)$ (Harrison
\cite{harrison74}); this effect is negligible for current redshift
catalogues. The first authors to arrive at this conclusion were
Nolthenius \& White (\cite{nolt87}); somehow, their careful analysis
has been forgotten lately. The situation is complicated yet by the
luminosity-density correlation -- brighter galaxies tend to be more
clustered, and the FoF scaling length should be calibrated by galaxy
brightness, also.

\begin{figure}[ht]
\centering
\resizebox{0.45\textwidth}{!}{\includegraphics*{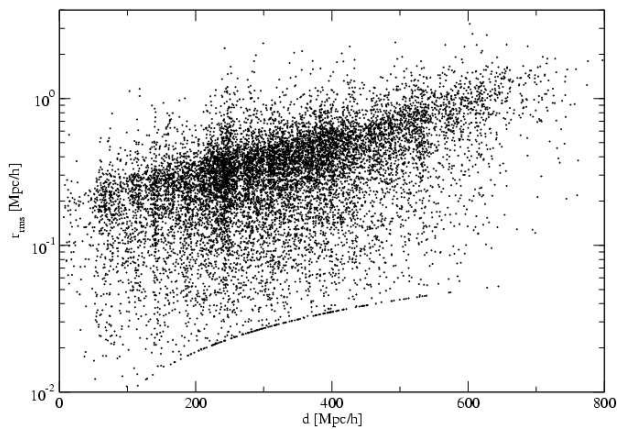}}
\caption{The rms sizes of groups of the 2dF GRS PIGG catalogue,
compiled by Eke \etal (\cite{eke04}) 
as a function of the comoving distance $d$ (for the Northern sample;
the Southern sample has a similar distribution).}
\label{fig:1}
\end{figure}

As the FoF scaling is rather complicated, all the resulting group
catalogues should be checked on the presence of systematic trends. For
the 2dFGRS, the only publicly available group catalogue is the 2PIGG
catalogue of Eke \etal (\cite{eke04}). They stress that the FoF
parameters should be chosen to guarantee that the group sizes should
not change while diluted by selection. This is really an important
requirement, otherwise our catalogue would consist of different
objects, those nearby and those far away.  Strangely, they do not
check if their choice of the FoF scaling satisfies this requirement,
although it is easy to do -- their catalogue includes the rms sizes of
the groups. We plot the rms sizes of the 2PIGG groups versus the group
distance in Fig.~\ref{fig:1}; as we see, a considerable trend remains.
We found a similar trend in a variable linking length FoF group
catalogue based on the SDSS DR1 (Einasto \etal \cite{ein04}).

\begin{figure*}[ht]
\centering
\resizebox{0.45\textwidth}{!}{\includegraphics*{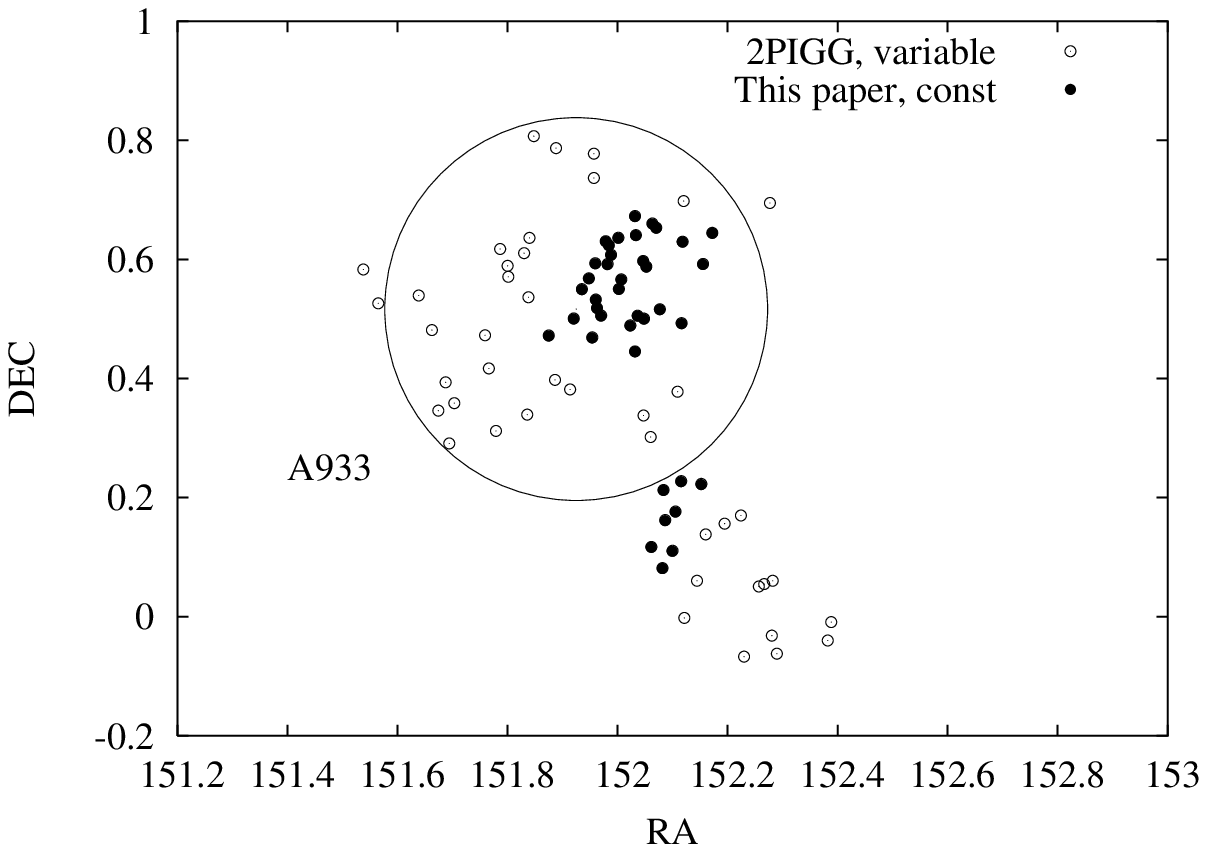}}\hspace{2mm} 
\resizebox{0.45\textwidth}{!}{\includegraphics*{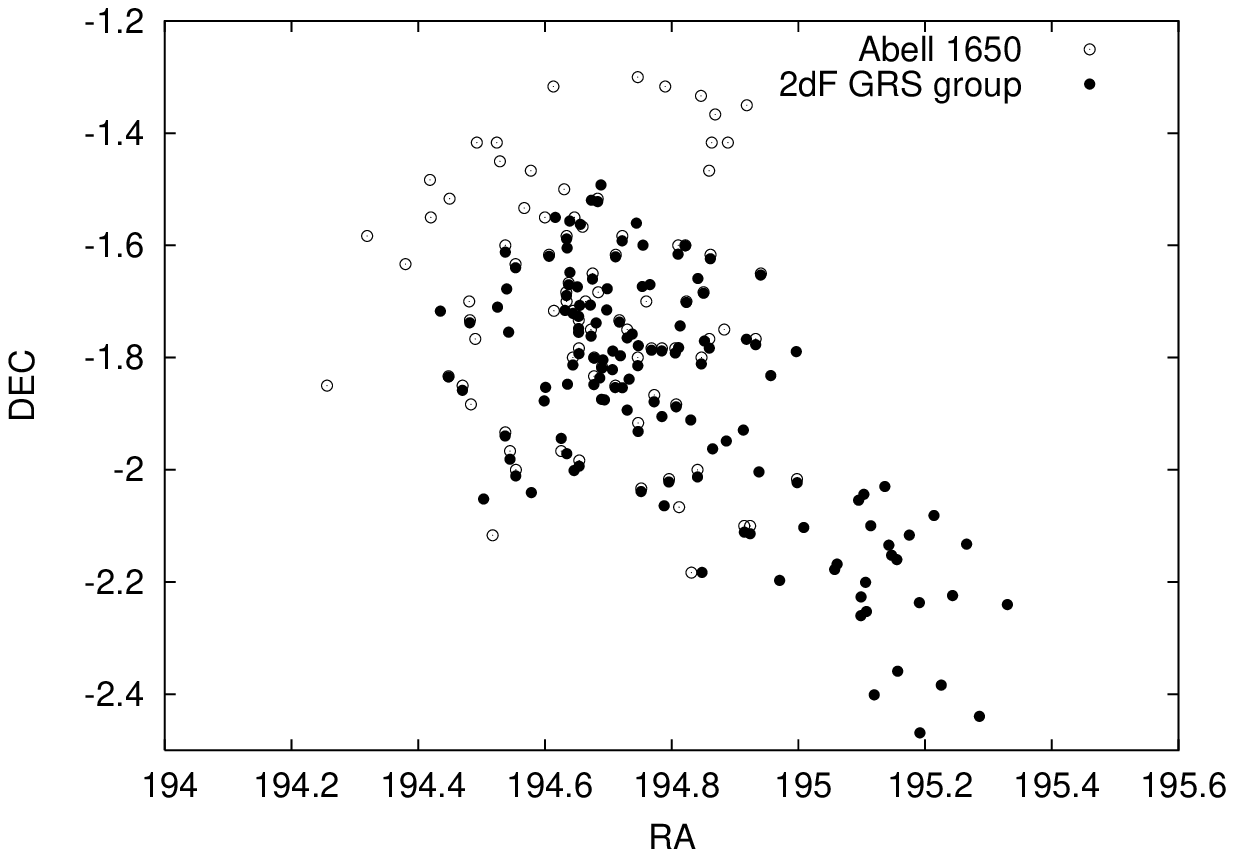}}\\
\caption{ The left panel compares the 2dF GRS groups obtained using the FoF method
with a constant linking length
(filled circles; two groups from our catalogue are shown)
and a variable linking length (open circles; a 2PIGG group) in the area of the
 cluster Abell 933 (in equatorial coordinates). The circle shows the
 size of the Abell cluster.  The right panel shows the distribution of
 galaxies in the cluster Abell 1650 and the corresponding 2dF GRS
 group, obtained without cleaning and using different linking lengths
 than in Table~\ref{Tab1} ($\Delta V$ = 500 km/s, $\Delta R
 $ = 0.5 Mpc).  }
\label{fig:2}
\end{figure*}

In order to study the effect of different FoF scalings, we generated
two versions of the group catalogue for the 2dF GRS, one with a
constant linking length, and another with a variable linking length
$l\sim f^{-1/3}$.  We compared the distribution of galaxies in rich
clusters using both versions of the catalogue, and the 2PiGG catalogue.  We also included for
comparison rich clusters by Abell (\cite{abell}) and Abell \etal
(\cite{aco}) as standard clusters, obtained using constant metric
radii for clusters.  The redshift data for the ACO clusters were taken
from the catalogue by Andernach \& Tago (\cite{at98}) and Andernach
\etal (\cite{and04}).  We present in Fig.~\ref{fig:2} two examples of
this comparison.  In the left panel of the figure we show the
distribution of galaxies in the sky for the rich galaxy cluster Abell
933.  In the 2PIGG catalogue (variable linking length), smaller clusters and a
number of nearby galaxies join to the cluster.
In our catalogue, in the case of constant linking length, the cluster includes only galaxies
inside the Abell radius of 1.5~\Mpc.  More distant galaxies are found
to form separate groups, which are less rich, and have smaller radii.
As another example we present in the right panel of Fig.~\ref{fig:2}
the FoF groups for the 2dF GRS in the vicinity of the cluster Abell
1650. We see that in the case of a too high linking length the 2dF GRS
group is more extended than the corresponding Abell cluster.

\begin{figure*}[ht]
\centering
\resizebox{0.45\textwidth}{!}{\includegraphics*{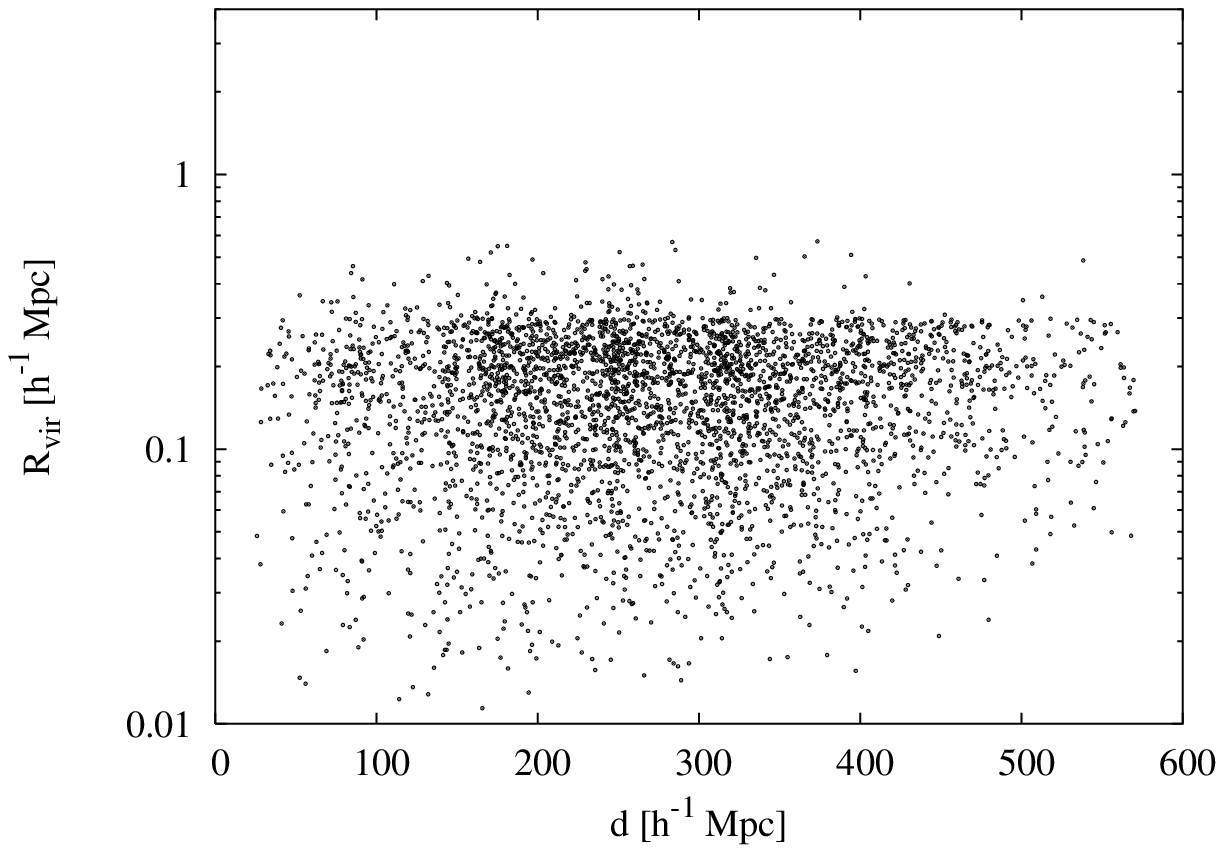}}
\hspace*{2mm}   
\resizebox{0.45\textwidth}{!}{\includegraphics*{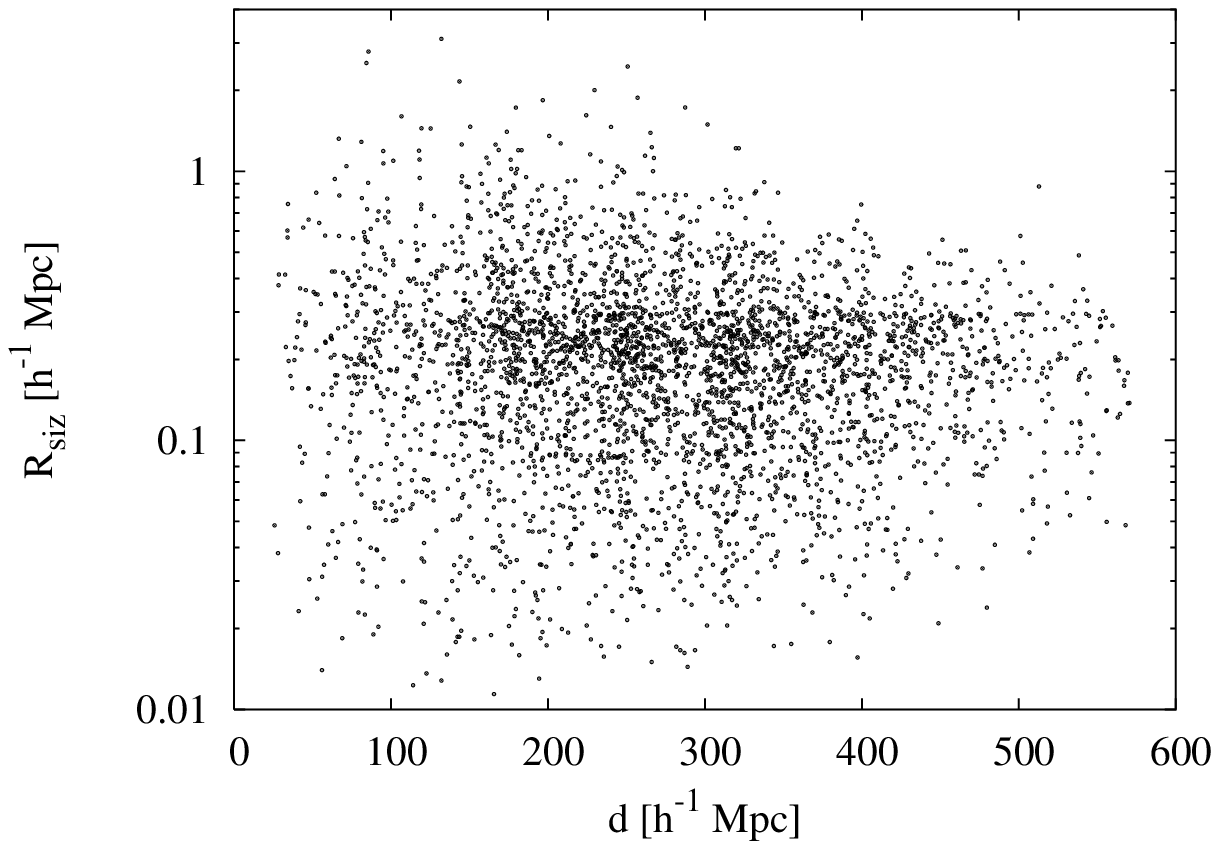}}\\
\caption{ 
  Sizes of our 2dF GRS groups, for a constant linking length. 
  Left panel -- the virial radius vs. distance, right panel --
  the maximum projected diameter vs. distance.}
\label{fig:3}
\end{figure*}

So, too high or increasing linking lengths cause considerable
extension of the groups and clusters, mainly by including nearby
filaments and surrounding smaller groups. This forces us to apply a
constant linking length in the FoF method, along with an additional
``cleaning'' procedure, described below.  In this case the mean virial
radii and the mean projected diameters of groups are practically
constant, see Fig.~\ref{fig:3}.  This figure confirms also that when
the virial radii are constant in the mean, then so are the maximum
diameters; we did not check it directly for the 2PIGG groups.

In order to find the best linking lengths, we found the groups, using
a number of different parameter values: $\Delta V$ = 100 - 900 km/s
and $ \Delta R $ = 0.2 - 1.0 \Mpc , and chose finally the values
presented in Table~\ref{Tab1}. These values of the FoF parameters
correspond to the best behaviour of group sizes with distance in
the 2dF GRS. Higher values for $\Delta R$ lead to inclusion of
galaxies from neighbouring groups and filaments. Lower values for
$\Delta V$ exclude the fastest members in intermediate richness
groups.

\begin{figure*}[ht]
\centering
\resizebox{0.45\textwidth}{!}{\includegraphics*{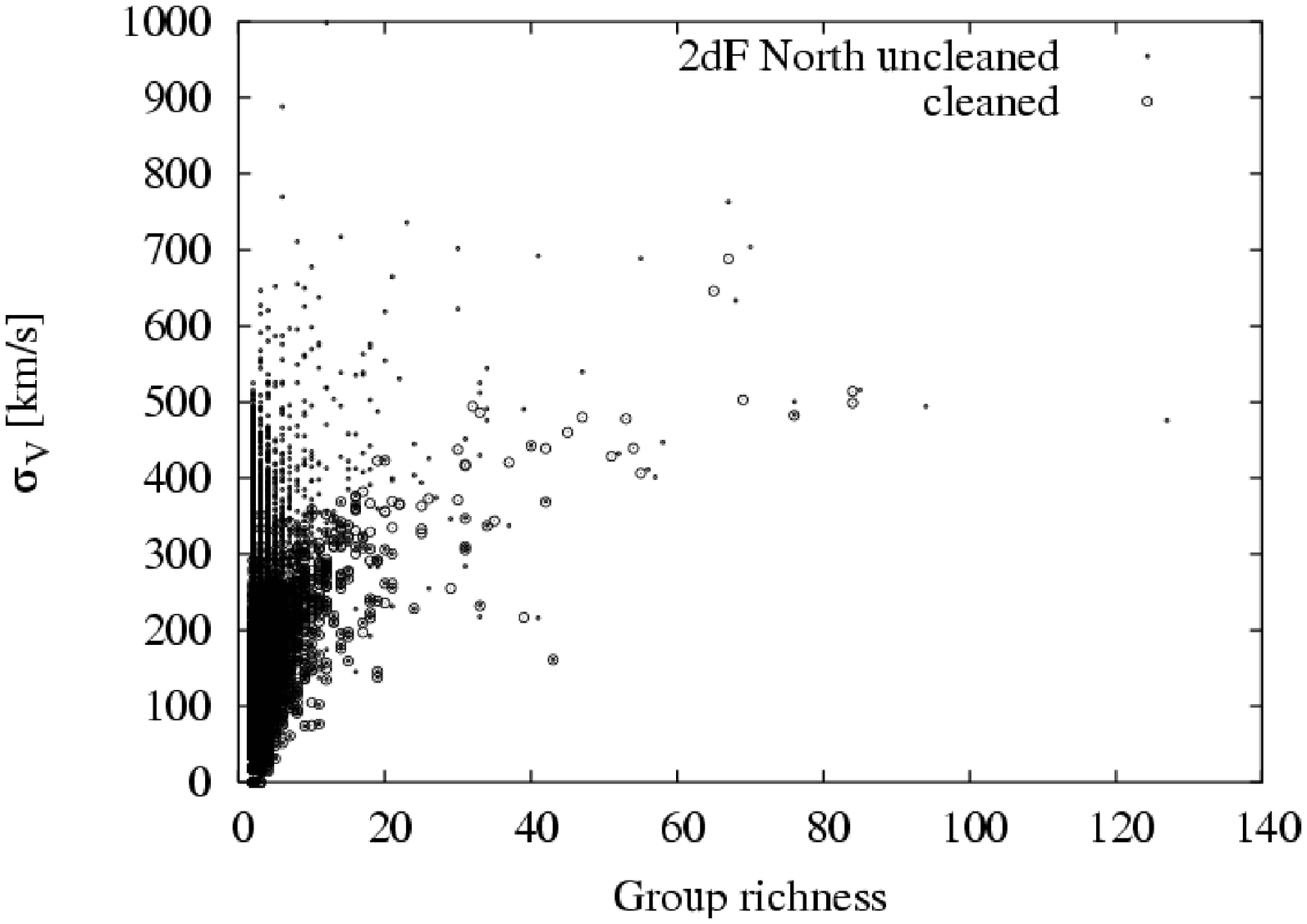}}
\hspace{2mm}
\resizebox{0.45\textwidth}{!}{\includegraphics*{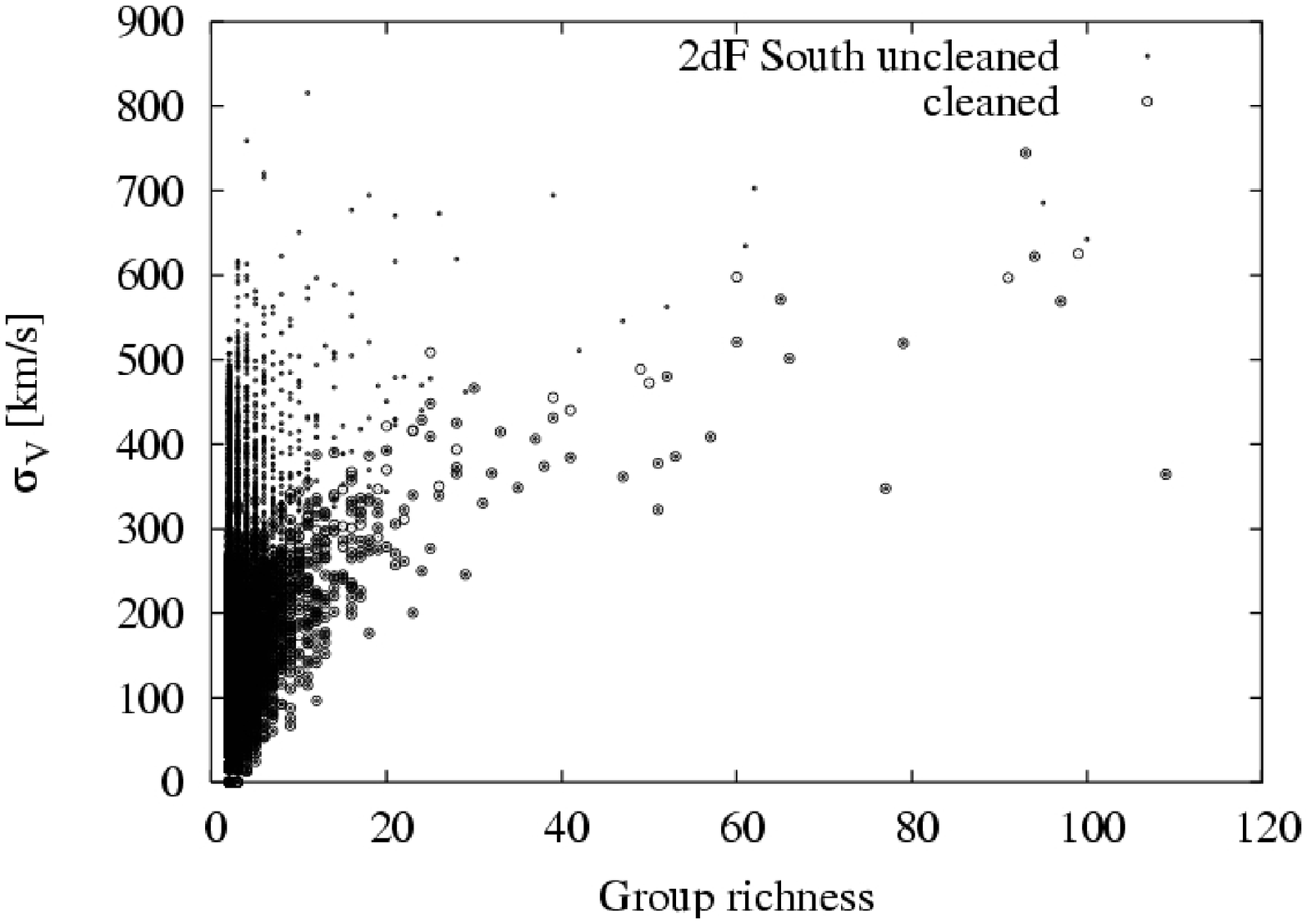}}\\
\caption{ Group rms velocities as a function of cluster richness
  for uncleaned (points) and cleaned (open circles) samples in
  the Northern (left panel) and the Southern patches (right panel).  }
\label{fig:4}
\end{figure*}

In addition, we applied a clean-up procedure to remove from groups
galaxies  with 
too high peculiar velocities. The procedure follows the
prescriptions proposed by Hein\"am\"aki \etal (\cite{hei03}), from
comparison of simulations and the LCRS groups. The essence of the clean-up
method is the comparison of the potential and kinetic energies of groups to
iteratively reject unbound members.  
This cleaning criterion can be expressed as:
\begin{equation}
\label{cleaning}
 (\sigma_V/V_0)^2 \geq N_g/(R_{vir} + R_0) 
\end{equation}
where $\sigma_V^2$ is the radial velocity dispersion , $V_0$ is a
calibrating parameter, $N_g$ is the number of group members, $R_{vir}$
is an estimate of the projected virial radius, and $R_0$ is a constant
which takes into account the difference between the spatial and
projected radii and has a value of a typical galaxy size (galaxies
can't be closer to each other than $R_0$). We know only the projected
value for $R_{vir}$, and estimate the radial velocity dispersion from
observations; therefore, the formula above has only statistical
meaning and may be in error for any particular group.

The groups found by the FoF method were checked for this criterion and
the rejected galaxies were checked for membership in the next step of
iteration; this allows us to break unbound groups into bound
subgroups. Rejection of the members for each particular group, which
do not satisfy the energy criterion, starts from the most distant
member from the group center in the velocity space and is iteratively
carried out until the criterion is satisfied or the whole group is
rejected. Then we start the procedure (FoF and the following clean-up)
again for the galaxies which were rejected as group members in the
previous iteration.  We found that three such global iteration steps
were enough, later steps yielded only bound groups and isolated
galaxies.

Fig.~\ref{fig:3} demonstrates that our method is free of distance
trends: the group virial radii do not
depend on redshift, in contrary to the 2PIGG groups by Eke et al.,
presented in Fig.~\ref{fig:1}.  
As we see in Fig.~\ref{fig:3}, the projected diametres of groups practically do
not change with redshift, being constant in the mean. One can observe a
scarcity of large systems after $d\approx400$\Mpc. 
This can be attributed to two effects: the decrease
of the multiplicity of groups (see Fig.~\ref{fig:8}) and the drop of
the total number of groups at that distance (see Fig.\ref{fig:7}).
At large distances only the brightest members of groups are visible,
and it is well known that bright galaxies tend to concentrate toward
the group centres.  Due to the decrease of the number of groups
statistisc may play a role: the probability of observing very large
(and very small) groups is low. 
This effect could also be partly caused by the
presence of rich superclusters at both the 2dF GRS patches at the
distances 200--400 \Mpc.

Fig.~\ref{fig:4} illustrates the
cleaning procedure, showing the rms velocity vs richness relation for
the groups before and after the clean-up. We see that this procedure
has excluded groups (mainly of low richness) which had too high
velocity dispersion, and has broken up a few high-richness groups.

The number of obtained groups and the FoF
parameters (for both equatorial slices) are given in Table~\ref{Tab1}.

\section{Group catalogue}

Our final catalogue includes 7657 Northern
and 10058 Southern groups with richness $\geq 2$. Both of the group
tables include the following columns for each group:

\begin{itemize}
 \item  identification number; 
 \item  richness (number of member galaxies); 
 \item  RA (J2000.0) in degrees (mean of member galaxies);
 \item  DEC (J2000.0) in degrees (mean of member galaxies); 
 \item  distance in \Mpc\ (mean of member galaxies); 
 \item  the maximum projected size (in \Mpc);
 \item  the rms radial velocity ($\sigma_V$, in km/s);  
 \item  the virial radius in \Mpc\ (the projected harmonic mean);
 \item  the luminosity of the cluster main galaxy (in units of $h^{-2}
 L_{\odot}$); 
 \item  the total observed luminosity of visible galaxies ($h^{-2}
 L_{\odot}$); 
 \item  the estimated total luminosity of the group  ($h^{-2} L_{\odot}$).
\end{itemize}

The identification number is attached to groups by the group finder in
the order the groups are found (independently for each of three
iterations, starting from 0, 15000 and 20000, respectively).
Calculation of luminosities is described in the next section.

We also present (in an electronic form) a catalogue of all individual
galaxies along with their group identification and the group richness,
ordered by the group identification number, to facilitate search.  The
tables of galaxies end with a list of isolated galaxies (or small
groups with only one bright galaxy within the observational window of
magnitudes); their group identification number is 0 and group richness
is 1.  All four tables can be obtained at
\texttt{http://www.aai.ee/$\sim$maret/2dfgr.html}.  Explanation of the
table columns is given in the file \texttt{2dfgroups.readme}.

\section{Luminosities of groups}

Galaxies were included in the 2dF GRS, if their corrected apparent
magnitude ${\rm b_j}$ lied in the interval from $m_1 = 13.5$ to $m_2 =
19.45$. The faint limit actually fluctuates from field to field, but
in the present context we shall ignore that; we shall take these
fluctuations into account in our future paper on the group luminosity
functions.

We regard every galaxy as a visible member of a group or cluster
within the visible range of absolute magnitudes, $M_1$ and $M_2$,
corresponding to the observational window of apparent magnitudes at
the distance of the galaxy.  To calculate total luminosities of groups
we have to find for all galaxies of the sample the estimated total
luminosity per one visible galaxy.  This estimated total luminosity
was calculated as follows (Einasto \etal \cite{e03b})
\begin{equation}
L_{tot} = L_{obs} W_L, 
\label{eq:ldens}
\end{equation}
where $L_{obs}=L_{\odot }10^{0.4\times (M_{\odot }-M)}$ is the
luminosity of a visible galaxy of an absolute magnitude $M$, and 
\begin{equation}
W_L =  {\frac{\int_0^\infty L \phi
(L)dL}{\int_{L_1}^{L_2} L \phi (L)dL}} 
\label{eq:weight2}
\end{equation}
is the luminous-density weight (the ratio of the expected total
luminosity to the expected luminosity in the visibility window).  In
the last equation $L_i=L_{\odot} 10^{0.4\times (M_{\odot }-M_i)}$ are
luminosity limits of the observational window corresponding to absolute
magnitude limits of the window $M_i$, and $M_{\odot }$ is the absolute
magnitude of the Sun.  In calculation of weights we assumed that galaxy
luminosities are distributed according to the Schechter
(\cite{S76}) luminosity function:
\begin{equation}
\phi (L) dL \propto (L/L^{*})^\alpha \exp {(-L/L^{*})}d(L/L^{*}),
\label{eq:schechter}
\end{equation}
where $\alpha $ and $L^{*}$ are parameters.  Instead of   $L^{*}$
the corresponding absolute magnitude $M^{*} - 5\log_{10} h$ is often 
used. 

\begin{figure}[ht]
\centering
\resizebox{0.45\textwidth}{!}
{\includegraphics*{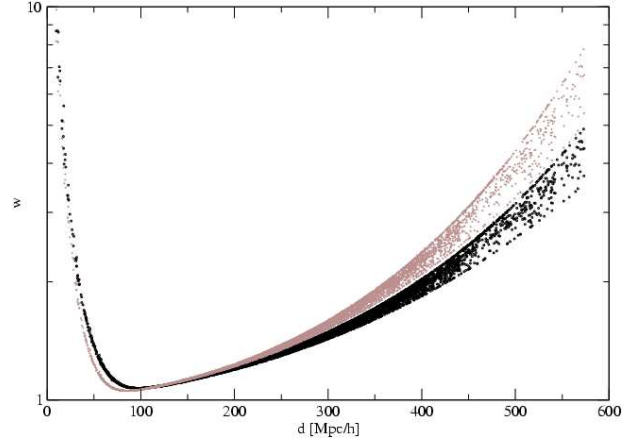}}
\caption{The mean weights of groups of the 2dF GRS Northern subsample as
  function of the distance from the observer.  Weights corresponding
  to Schechter parameter set 1 are plotted by gray symbols, weights
  for parameter set 2 with black symbols.
}
\label{fig:5}
\end{figure}

Following Eke \etal (\cite{eke04}) we accepted $M_{\odot} = 5.33$ in
the ${\rm b_j}$ photometric system.  Further we have adopted the $k +
e$-correction according to Norberg \etal (\cite{2002MNRAS.336..907N})
as follows: morphological Type 1: $k + e = (2z +2.8z^2)/(1+3.8z^3)$;
Type 2: $k + e = (0.6z + 2.8z^2)/(1 + 19.6z^3)$; Type 3: $k + e = (z +
3.6z^2)/(1 + 16.6z^3)$; Type 4: $k + e = (1.6z + 3.2z^2)/(1 +
14.6z^3)$.  The morphological types were defined as follows: Type 1:
$\eta < -1.4$; Type 2: $-1.4 \leq \eta < 1.1$; Type 3: $1.1 \leq \eta
< 3.5$; Type 4: $3.5 \leq \eta$, where $\eta$ is the spectral
classification parameter, given in the 2dF GRS dataset.  For some
galaxies with poor spectra the spectral type parameter $\eta$ is not
determined; in these cases we applied the mean relation: $k + e = (z +
3.2z^2)/(1 + 20z^3)$ (N02).

We used two sets of Schechter parameters to calculate the luminosity
weights: $\alpha_1 = -1.21$, $M^{*}_1 - 5\log_{10} h = -19.66$, and
$\alpha_2 = -1.28$, $M^{*}_2 - 5\log_{10} h = -20.07$.  The first set
was found by N02 for the whole 2dF galaxy sample, the second set by De
Propris et al. (\cite{dep02b}) for cluster galaxies.  We calculated
for each group the total observed and corrected luminosities,
and the mean weight
\begin{equation}
W_m = {\frac{\sum L_{tot,i}} {{\sum L_{obs,i}}}},
\label{eq:sum}
\end{equation}
where the subscript $i$ denotes values for individual observed galaxies in
the group, and the sum includes all member galaxies of the system.

The mean weights for the groups of the 2dF GRS Northern subsample are
plotted as a function of the distance $d$ from the observer in
Fig.~\ref{fig:5}. We see that the mean weight has a value a bit higher
than unity at distance $d\sim 80$~\Mpc, and increases both toward
smaller and larger distance.  The increase at small distances is due
to the absence of very bright members of groups, which lie outside the
observational window, and at large distance due to the absence of faint
galaxies. The weights grow fast for very close groups and for groups
farther away than about 400\Mpc. At these distances correction factors
start to dominate and the luminosities of groups become uncertain.

\begin{figure*}[ht]
\centering
\resizebox{0.45\textwidth}{!}
{\includegraphics*{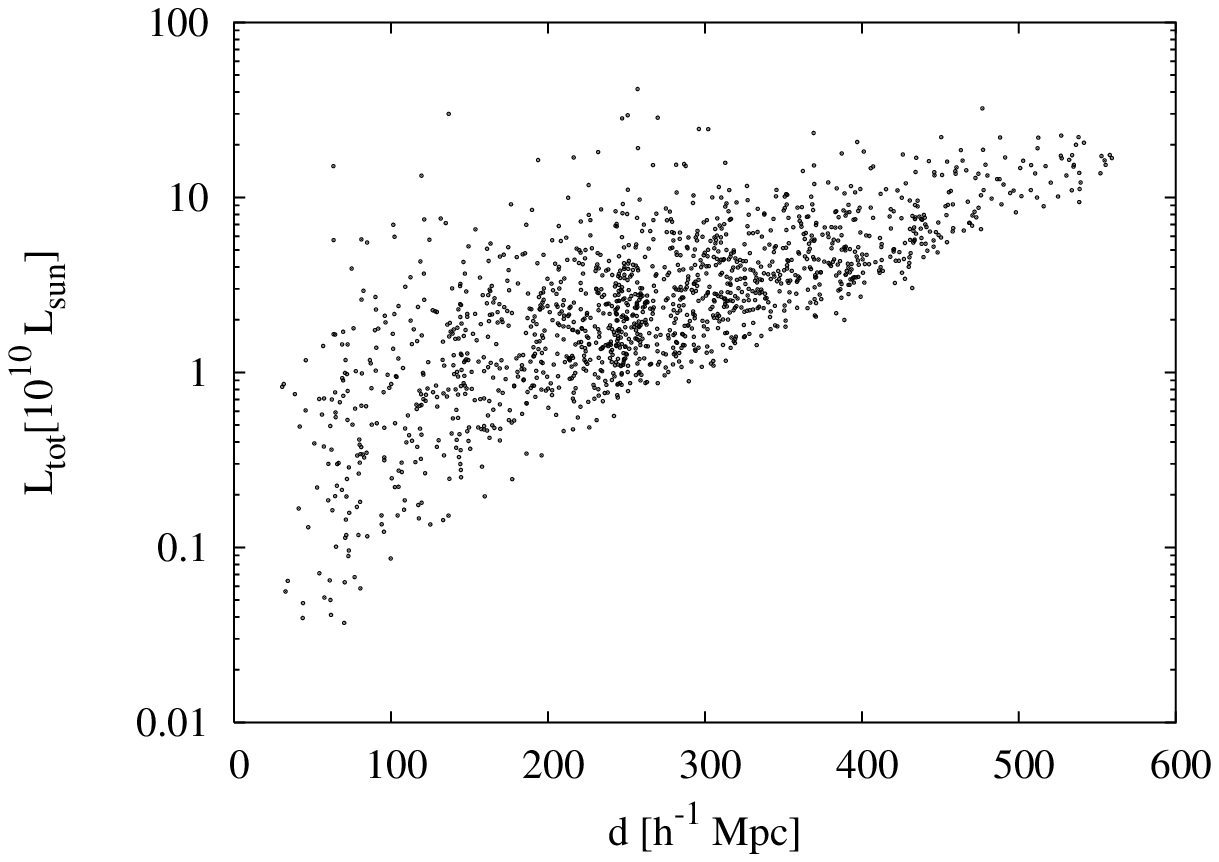}}\hspace{2mm}
\resizebox{0.45\textwidth}{!}
{\includegraphics*{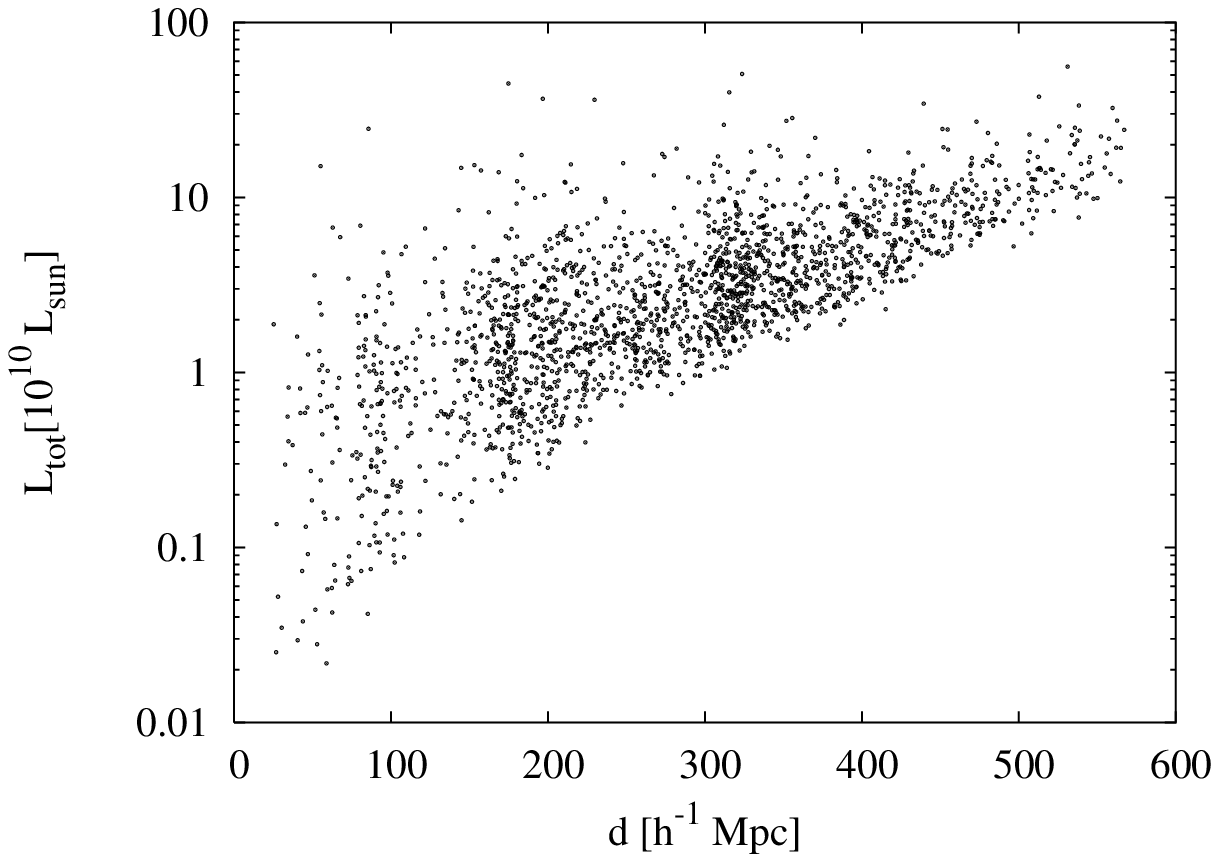}}\hspace{2mm} 
\caption{The estimated total luminosities of clusters of
  galaxies of the 2dF at various distance from the
  observer. The left panel shows groups/clusters of the Northern
  section, the right panel of the Southern section.
}
\label{fig:6}
\end{figure*}

We also see that at large distances the weights corresponding to the
Schechter parameter set 1 are systematically higher than the weights
of the set 2.  This difference is due to the fact that the Schechter
function, calculated for the whole galaxy sample, takes into account
also the presence of groups which have no galaxies within the
observational window (i.e. all galaxies of the group are too faint);
the luminosity of these galaxies is tacitly added to the luminosity of
visible groups (for a detailed discussion of this selection effect see
Einasto \etal \cite{e03b}).  The parameter set 2 was determined using
only galaxies, which lie in visible groups and clusters; thus this set
should yield unbiased values for weights and total luminosities.  In
the following we have used only this set of weights and luminosities.

\begin{figure*}[ht]
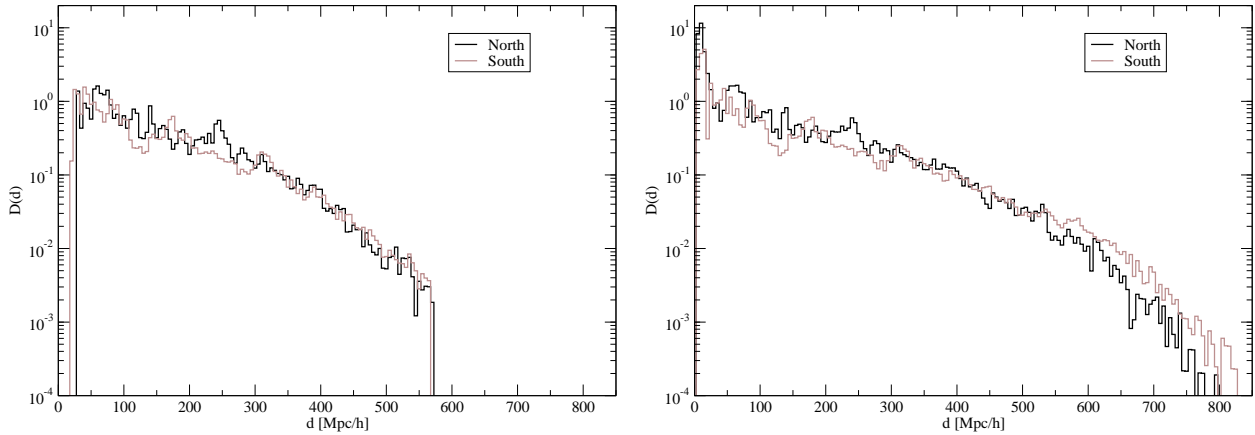

\centering
\resizebox{0.45\textwidth}{!}{\includegraphics*{erik_fig7a.eps}}
\hspace{2mm}
\resizebox{0.45\textwidth}{!}{\includegraphics*{erik_fig7b.eps}}
\hspace{2mm} 
\caption{The spatial density of groups as a function of
  distance from the observer.  The left panel shows data for the groups
  found in the present paper, the right panel for the 2PIGG groups.
}
\label{fig:7}
\end{figure*}

In Fig.~\ref{fig:6} we show the estimated total luminosities of groups
as a function of distance.  We produced also colour figures that
visualize the luminosities of groups. These are too detailed to
present here, and can be found at our web pages.  These figures show
that the brightest groups have corrected total luminosities, which
are, in the mean, independent of distance.  This shows that our
calculation of total luminosities is correct.  

\section{Selection effects}

The main selection effect in surveys like the 2dF GRS is caused by a
fixed interval of apparent magnitudes, $m_1 \dots m_2$, used in
selection of galaxies for redshift measurements.  This interval
translates to an absolute magnitude interval, $M_1 \dots M_2$, that
depends on the distance from the observer $d$.  This effect is well
seen in Fig.~\ref{fig:6}, where the estimated total luminosities of
our groups/clusters are shown as a function of the distance $d$.  We
see a well-defined lower luminosity limit of groups that increases
with distance.  This low-luminosity limit is due to the faint-end
limit of the 2dF observations, $m_2\approx 19.45$ in ${\rm b_j}$.

\begin{figure*}[ht]
\centering
\resizebox{0.45\textwidth}{!}{\includegraphics*{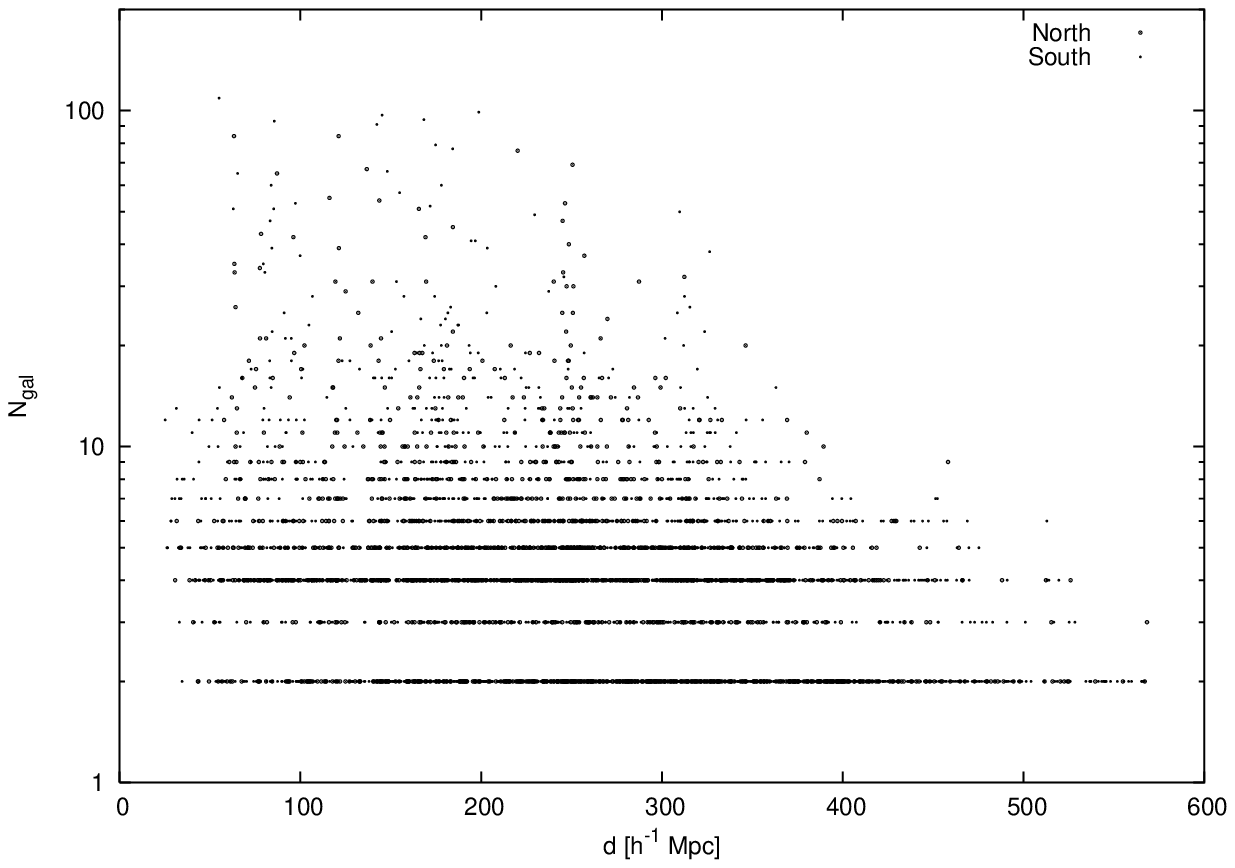}}
\hspace{2mm}
\resizebox{0.45\textwidth}{!}{\includegraphics*{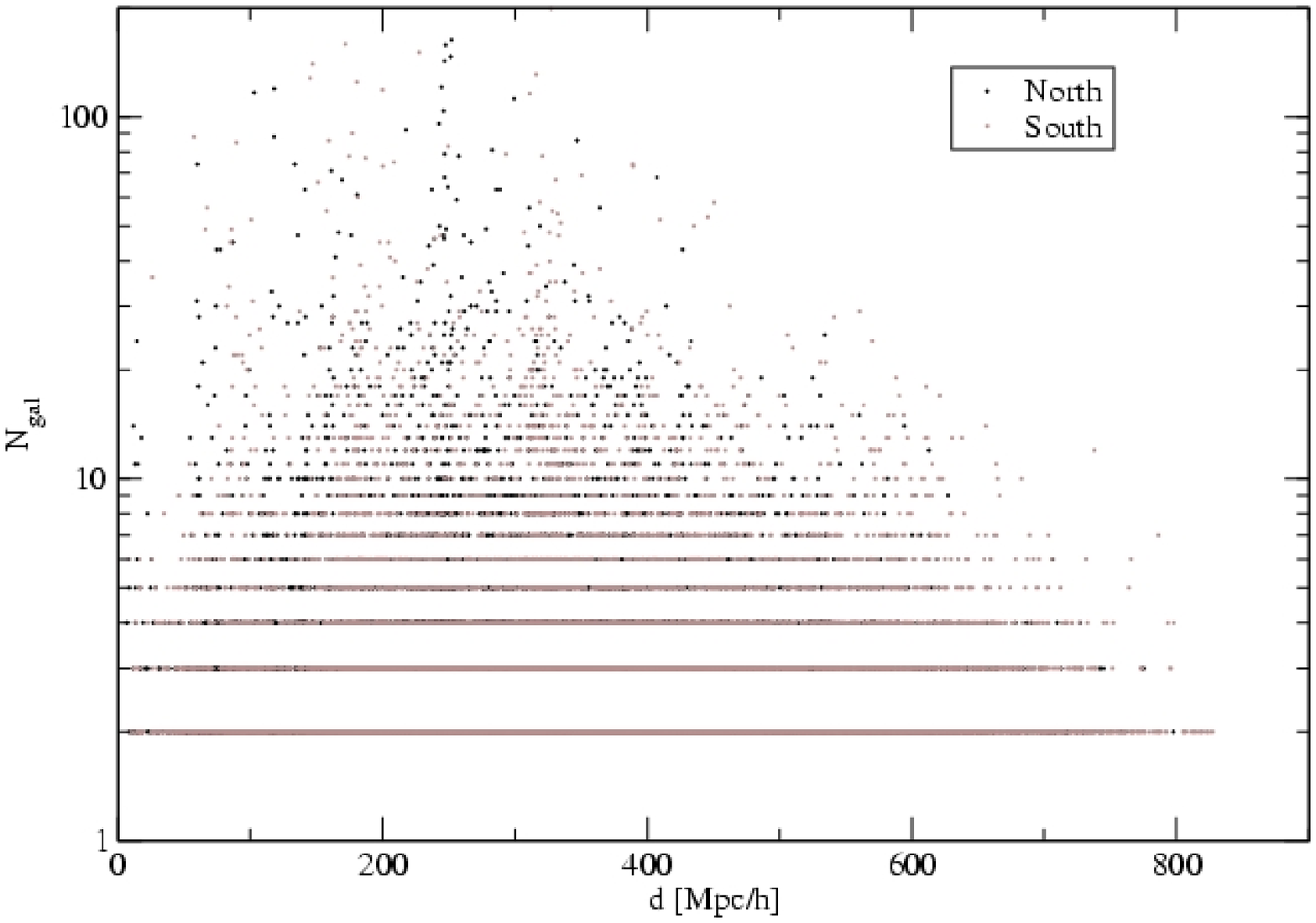}}
\hspace{2mm} 
\caption{The multiplicity of groups as a function of
  distance from the observer.  The left panel shows data for the groups
  found in the present paper, the right panel for the Eke groups.
}
\label{fig:8}
\end{figure*}

This selection effect affects groups in several ways.  First
of all, in nearby space our group-finding algorithm finds faint as
well bright groups, whereas at large distance only bright groups can
be found. This is the effect shown in Fig.~\ref{fig:6}.   

The primary consequence of this selection effect is the decrease of
the volume density of groups at large distances.  The mean volume
density of groups in shells of fixed thickness as a function of
distance is plotted in Fig.~\ref{fig:7}, separately for the Northern
and Southern strip of the 2dF GRS.  For comparison we give a similar
distribution also for the groups of the Eke catalogue.  We see that
the spatial density decreases by a factor of about 1000 for the
present catalogue.  The Eke catalogue has greater depth and includes a
number of very faint groups in the nearby volume; for this reason the
spatial density interval is even larger, almost 5 orders of
magnitude. This decrease must be taken into account in statistical
analysis of catalogues, usually by selection functions. As the
correction amplitudes are large, the results of statistical analysis
of group catalogues can easily become correction-dominated.

{\scriptsize
\begin{table*}
      \caption[]{Data on groups from catalogues based on the 2dF GRS}
         \label{Tab2}
      \[
         \begin{tabular}{lccccclcc}
            \hline\hline
            \noalign{\smallskip}
            Sample & $N_{gal} $ &
      	$N_{gr}(n \geq 2)$&$N_{gr}(n \geq 4)$&$z_{lim}$ & $m_{lim}$ &
	method & \% in clusters  & available \\ 
            \noalign{\smallskip}
            \hline
            \noalign{\smallskip}

Merchan 2002 & 60000 &  -   & 2209  & 0.25 & 19.45 & modified FoF &  & no\\

Eke 2004 & 191440   & 28877 & 7020  & 0.11 (median) &   &  FoF with
test mock & 55 \% & yes\\ 

Yang 2004 & 150715  & 11434 & 2471  & 0.2  & 19.45  & FoF modified +
mock test & 60 \%  & no\\  

Tago 2004 & 184495  & 17715 & 3044  & 0.009 - 0.2  &      & FoF +
clean-up  & 30 \% & yes\\  

            \noalign{\smallskip}
            \hline
         \end{tabular}
      \]
   \end{table*}
}

The limited magnitude interval introduces one more selection effect:
it decreases the number of galaxies seen in groups.  The
low-luminosity end of the observational window of absolute magnitudes
shifts with increasing distance toward brighter magnitudes.  As a
result, faint members of groups fall outside the observational limit,
and the multiplicity of the group decreases.  In Fig.~\ref{fig:8} we
show the multiplicity of groups (the number of galaxies) as a function
of distance from the observer, separately for Northern and Southern
groups.  A similar plot is given for the groups by Eke
et al., for comparison. We see that rich groups are seen only up to a
distance of about 300~\Mpc, thereafter the mean multiplicity decreases
considerably with distance.  This selection effect must be accounted
for in multiplicity analysis.

\begin{figure}[ht]
\centering
\resizebox{0.50\textwidth}{!}{\includegraphics*{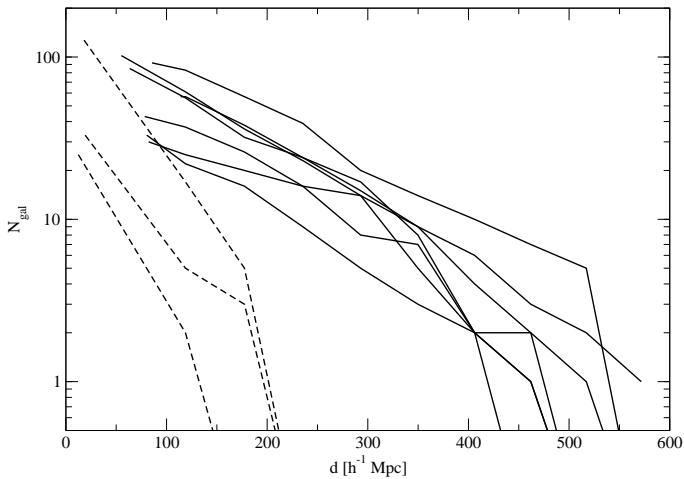}}
\caption{The number of galaxies in groups, shifted progressively to
  a larger distance $d$.  Solid lines show the results for groups
  with intrinsically bright galaxies, dashed lines for groups with
  fainter member galaxies. 
}
\label{fig:9}
\end{figure}

To illustrate the selection effect on cluster multiplicity, we made
the following test.  We selected in the nearby volume ($d < 100$~\Mpc)
a number of groups with multiplicity $N_{gal} > 20$, shifted the
groups progressively to larger distances, and calculated the
multiplicity of the group, if it were located at this distance. As
with increasing distance more and more fainter members of groups fall
outside the observational window of apparent magnitudes, the
multiplicity decreases with distance.  The results of our calculations
for ten groups are shown in Fig.~\ref{fig:9}.  We see that the number
of galaxies within the observational window decreases in logarithmic
scale almost linearly.  The slope of this power law depends on the
luminosity function of galaxies in groups.  Groups populated mainly
with faint galaxies lose their members more rapidly than groups
populated with brighter galaxies, as expected.  The majority of groups
selected for this test have at a certain distance only one visible
galaxy left, that becomes also invisible if the distance increases
further.  At the largest distance only one group selected for this
test is visible as a single galaxy, all other groups have become too
faint to be detected within the visibility window of the 2dF GRS.

\section{Discussion and conclusions}

\subsection{Comparison to other studies}

Here we compare various group catalogues obtained for the 2dF GRS.  We
have found five papers on groups, based on the 2dF GRS galaxies.  The
first attempt to study the 2dF GRS clusters of galaxies, based on
well-known previous clusters, was done by De Propris \etal
(\cite{dep02a}).  However, as their catalogue starts from the known
clusters (Abell \cite{abell}; Abell \etal \cite{aco}; APM; EDCC) and
not from the 2dF GRS sample, we have not included it in our comparison
of the 2dF GRS groups.  Table~\ref{Tab2} presents data for groups from
various catalogues, based on the 2dF GRS.

Several studies have shown (see, e.g., Kim \etal \cite{kim02}) that
different methods give rather different groups for the SDSS
sample. The same is true for the 2dF GRS groups.  Although all four
methods in Table~\ref{Tab2} are FoF-based, the results are remarkably
different.  In particular, Yang \etal (\cite{yang04}) applied a more
elaborate group finder.  They noted that the number of clusters they
found corresponds to numbers obtained by Eke et al., concluding,
however that their cluster finder is better than that used by Eke et
al. Unfortunately, the groups by Yang \etal (\cite{yang04}) are not
publicly available, and we cannot compare them with our groups.

Single galaxies can be considered as belonging to small groups or to
halos represented by one observed galaxy.  The fraction of galaxies in
groups by Eke et al., Yang \etal and in this paper is 55, 60 and 30\%,
respectively (these are the catalogues where the data are available).
Different fractions of galaxies in groups (or the fractions of single
galaxies) lead to the different number density of halos, as well as to
different distributions of the halo occupation number.  A very
important problem of substructure was stressed and extensively studied
by Burgett \etal (\cite{bur04}).  Their conclusion is that most of the
clusters have substructures which merge into a large cluster in a very
long dynamical time scale. Therefore, we are faced with the problem,
if we have to count a collection of different haloes in process of
merging as one cluster or a number of sub-clusters.  The large number
of merging (sub)groups (and the existence of a large number of groups
of low richness) is a justification for a large fraction of single
galaxies in our results.  Our group finder is tuned for more compact,
and gravitationally bound groups

\subsection{Conclusions}

The main results of the present paper are following:

\begin{itemize}

\item{} we present a catalogue of groups and clusters of galaxies of the 2dF
GRS final data release. We applied the FoF method with a constant
linking length, in which case the properties of groups depend less on
their redshift, than in the conventional case of a variable linking
length;

\item{} we applied a clean-up procedure to our groups that compares
the potential and kinetic energies of groups, to obtain
a statistically reasonable catalogue of bound groups;

\item{} we analyzed selection effects in group catalogues and
  compared of our group catalogue to the catalogue  
by Eke \etal (\cite{eke04}); we found that the catalogues differ
for group densities, multiplicities and sizes.

\item{} we present our catalogue at our web page
(\texttt{http://www.aai.ee/$\sim$maret/2dfgr.html}); we hope
that it will serve as a basis for further studies of large-scale structure.

\end{itemize}

\begin{acknowledgements}

   We are pleased to thank the 2dF GRS Team for the publicly available 
   final data release.  We acknowledge the Estonian Science Foundation for
   support from Grant No. 4695.

\end{acknowledgements}


\begin{thebibliography}{}

\bibitem[1958]{abell} Abell, G. 1958, ApJS, 3, 211

\bibitem[1989]{aco} Abell, G., Corwin, H., \& Olowin, R. 1989, ApJS,
70, 1

\bibitem[1998]{at98} Andernach, H., \& Tago, E. 1998, in {\em Large
Scale Structure: Tracks and Traces}, eds.\ V.~M\"uller,
S.\,Gottl\"ober, J.P.\,M\"ucket \& J.\,Wambsganss, World Scientific,
Singapore, p.\ 147

\bibitem[2004]{and04} Andernach, H., Tago, E., Einasto, M., Einasto, J., 
Jaaniste, J. 2004,
In "Nearby Large-Scale Structures and the Zone of Avoidance",
eds. A.P. Fairall, P. Woudt, ASP Conf. Series, San Francisco:
Astronomical Society of the Pacific, (astro-ph/0407097)

\bibitem[2003]{bac03} Bahcall, N., McKay, T.A., Annis, J.,
        \etal 2003,  ApJS, 148, 243

\bibitem[2004]{bur04} Burgett, M., Vick, M.M., Davis, D.S., et al.
  2004, (astro-ph/0405021) 

\bibitem[2001]{car01} Carlberg, R. G., Yee, H. K. C., Morris, S. L., \etal 
       2001 , ApJ, 552, 427

\bibitem[2001]{col01} Colless, M.M., Dalton, G.B., Maddox, S.J., \etal
     2001, MNRAS, 328,  1039, (astro-ph/0106498)

\bibitem[2003]{col03} Colless, M.M., Peterson, B.A., Jackson, C.A., \etal 2003,
         (astro-ph/0306581) 

\bibitem[2002a]{dep02a} De Propris, R., \etal (2dF GRS Team) 2002a,
  MNRAS, 329, 87 

\bibitem[2002b]{dep02b} De Propris, R., \etal (2dF GRS Team) 2002b,
  MNRAS, 342, 725, (astro-ph/0212562)

\bibitem[2003a]{e03a} Einasto, J., Einasto, M., H\"utsi, G., \etal
   2003a,  A\&A, 410, 425

\bibitem[2003b]{e03b} Einasto, J.,  H\"utsi, G., Einasto, M., \etal~
     2003b,  A\&A, 405, 425

\bibitem[1984]{e84} Einasto, J., Klypin, A.A., Saar, E. \& Shandarin,
  S. F. 1984, MNRAS, 206, 529

\bibitem[2004a]{ein04}  Einasto, J., Tago E., Einasto, M., 
     Saar, E. 2004a,
In "Nearby Large-Scale Structures and the Zone of Avoidance",
eds. A.P. Fairall, P. Woudt, ASP Conf. Series, San Francisco:
Astronomical Society of the Pacific, in press, (astro-ph/0408463) 


\bibitem[2001]{e2001} Einasto, M., Einasto, J., Tago, E., M\"uller, V.,
            \& Andernach, H.  2001, AJ, 122, 2222 (E01)

\bibitem[2004a]{einm04} Einasto, M., Suhhonenko, I., Hein\"am\"aki,
    P., Einasto, J., \& Saar, E.  2004a, submitted to A\&A, (astro-ph/0411529) 

\bibitem[2004]{eke04} Eke, V. R., Baugh, C. M., Cole, S., et al.
  2004, (astro-ph/0402567)

\bibitem[2002]{gal02}    Gal, R., Lopes, P. A. A., Djorgovski, S. G., \etal
    (DPOSS Team) 2002, A\&AS, 200. 9711

\bibitem[2003]{gal03}    Gal, R. R., de Carvalho, R. R., Lopes, P. A. A., \etal
     2003, AJ, 125, 2064

\bibitem[2004]{ger04} Gerke, B., Newman J., Davis0 M., \etal 2004,
  (astro-ph/0410721) 

\bibitem[2001]{goto} Goto, T.,  Sekiguchi, M., Nichol, R.C., 
  for the SDSS collaboration, 2001, ApJ, in press, (astro-ph/0112482) 

\bibitem[1974]{harrison74} Harrison, E.R. 1974, ApJL, 191, 51

\bibitem[2003]{hei03} Hein\"am\"aki, P., Einasto, J., Einasto, M., \etal~
  2003, A\&A, 397, 63

\bibitem[1982]{hg82} Huchra, J. P., \& Geller, M. J. 1882, ApJ, 257, 423

\bibitem[2002]{kim02} Kim, R.S.J., Kepner, J.V., Postman, M., \etal~
   2002, AJ, 123,  20, (astro-ph/0110259)

\bibitem[2004]{lee04} Lee, B.C., \etal 2004, AJ, 127, 1811

\bibitem[2004]{lop} Lopes, P.A.A., de Carvalho, R.R., Gal, R.R., \etal
   2004, (astro-ph/0406123)


\bibitem[2003]{mar03} Mart{\'\i}nez, V.J., \& Saar, E. 2003. Statistics
  of the Galaxy Distribution. Chapman Hall/CRC, Boca Raton, 432 pp.

\bibitem[2002]{mer03} Merchan, M., \& Zandivarez, A. 2002, MNRAS, 335, 216. 

\bibitem[2004]{mer04} Merchan, M., \& Zandivarez, A. 2004, (astro-ph/0412257)

\bibitem[2004]{nic04} Nichol, R. 2004, Carnegie Obs. Astroph. ser.,
  p.24, (astro-ph/0305041)  

\bibitem[1987]{nolt87} Nolthenius, R., White, S.D.M. 1987, MNRAS, 235, 505

\bibitem[2002]{2002MNRAS.336..907N} Norberg, P., et al. 2002, MNRAS, 336,
907,  (astro-ph/0111011)

\bibitem[1976]{S76} Schechter, P. 1976, \apj, 203,
	297

\bibitem[2000]{Tucker00} Tucker, D.L., Oemler, A.Jr., Hashimoto, Y., \etal
 2000, ApJS, 130, 237

\bibitem[2004]{yang04} Yang, X., Mo, H.J., van den Bosch, F.C.,\&
  Jing, Y.P. 2004 , (astro-ph/0405234)

\bibitem[1982]{zes82}  Zeldovich, Ya.B., Einasto, J., \&
Shandarin, S.F. 1982, Nature, 300, 407

\end{thebibliography}
\end{document}